# A Timing Yield Model for SRAM Cells in Sub/Near-threshold Voltages Based on A Compact Drain Current Model

Shan Shen, Peng Cao, Ming Ling, and Longxing Shi

***Abstract*—Sub/Near-threshold static random-access memory (SRAM) design is crucial for addressing the memory bottleneck in energy-constrained applications. However, the high integration density and reliability under process variations demand an accurate estimation of extremely small failure probabilities. To capture such a "rare event" in memory circuits, the time and storage overhead of conventional Monte Carlo (MC) simulations cannot be tolerated. On the other hand, classic analytical methods predicting failure probabilities from a physical expression become inaccurate in the sub/near-threshold region due to the assumed distribution or the oversimplified drain current model for nanoscale devices. This work first proposes a simple but efficient drain current model to describe the drain-induced barrier lowering effect at low voltages. Based on that, the probability density functions of the interest metrics in SRAM are derived. Two analytical models are then put forward to evaluate SRAM dynamic stabilities including access and write-time failures. The proposed models can be extended easily to different types of SRAM with different read/write assist circuits. The models are validated against MC simulations across different operating voltages and temperatures. The average relative errors at 0.5V $V_{DD}$ are only 8.8% and 10.4% for the access-time and write failure models respectively. The size of required data samples is 43.6X smaller than that of the state-of-the-art method.**

## I. Introduction

Memory cell circuits need to be replicated millions of times to achieve extremely high integration density in a small footprint, where cutting-edge process technology is demanded. In this case, the stringent yield requirement of memory circuits can be translated into an extremely small failure probability of each component circuit, thereby making the circuit failure a "rare event". In general, the failure probability estimation is usually analytically intractable due to the high complexity of memory circuits. The most straightforward approach is the Monte Carlo (MC) method, which repeatedly draws samples and evaluates circuit performance by transistor-level SPICE simulations. However, MC is extremely time-consuming for the rare-event estimation, because millions or even billions of samples are needed to capture one failure.

Several analytical models [1][2][3] are proposed to solve this problem, while work [4] uses Gaussian-tail-fitting to model the distribution of one-sided static noise margin. Albeit they achieve great accuracy at nominal voltages, predictions deviate from simulation results at low voltages and low failure probabilities [5]. This is because (1) the cell read current or differential voltage on bit-lines becomes non-Gaussian at low supply voltages, and (2) the drain current ($I_{ds}$) function from which analytical models are derived is either unsuitable for transregional near-threshold voltages or ignores the short-channel effect. The analytical methods based on static noise margin (SNM) [6][7] and dynamic noise margin (DNM) [8] are proposed to help the bitcell design, but these metrics only account for read/write margins to balance the sizes of transistors, rather than timing violations. Voltage Acceleration Method (VAM) [9] measures read current ($I_{READ}$) at the high-sigma region from actual silicon by aggregating the $I_{READ}$ distributions at lower voltages. Besides $I_{READ}$, the functional failures of SRAM are also affected by variations of peripheral circuits and the length of wordline (WL) pulse [3].

Furthermore, the SRAM frequency depends on the longest path of the access and write operations. As a result, we need to consider both in the failure analyses. Other types of failure models, such as those for static timing analysis (STA) [10][11], can be also used by SRAM to model the write failures because it leverages 2 cross-coupled inverters to store data. But the yield requirement of a logic path is much looser than that of a memory cell.

Other researchers use importance sampling to largely reduce the number of samples required by MC simulations. The Adaptive Importance Sampling method (AIS) [12] develops an iterative resampling approach to search for failure regions, which can further reduce the simulation samples. High Dimensions Importance Sampling (HDIS) [13] is proposed to solve the high dimension problem of process parameters when the scale of circuits is large. Unfortunately, the importance-sampling-based methods have 2 main problems. Unlike digital circuits that can be efficiently analyzed at the gate level, SRAM including most analog/mixed-signal circuits must be simulated at the transistor level. The time of a single simulation is still large. Although these types of analysis methods can decrease the number of simulations, they still require tens of thousands

Shan Shen, Peng Cao (corresponding author), Ming Ling, and Longxing Shi are all with the Nation ASIC System Engineering Technology Research Center, Southeast University, Nanjing, 210096, China (e-mail: caopeng@seu.edu.cn).

of training samples to get a converged result compared to analytical yield models. On the other hand, the prediction results of these methods may have large fluctuations at some conditions [5], which is because the likelihood ratios between the original sampling distribution and the distorted sampling one have huge numerical instability [34].

Cell yield analyses using Low-Rank Tensor Approximation (LRTA) [14] and Spline-High Dimensional Model Representation (SP-HDMR) [15] substitute the expensive simulations in failure probability estimation for meta-models and integrate their models into AIS and HDIS, respectively. These methods are equivalent to trade the estimation precision for the simulation speedup because the accuracy of meta-models greatly affects the final predicted results. Besides, constructing and solving the meta-models are still complicated.

In a word, accuracy and speedup are a tradeoff in failure analyses, where analytical models are efficient enough but suffer from poor accuracy, particularly at low supply voltages, while sampling-based methods improve the precision but cost lots of samplings or computation resources. Thus, in this paper, our goal is to probe the boundary of the trade-off in the failure probability estimation of SRAM cells. Our analysis starts with a compact $I_{ds}$ model. Then, analytical models for evaluating access-time and write failure probabilities are constructed. Our contributions include the following.

- We improve the accuracy of a compact drain current model for the near/sub-threshold region by considering drain induced barrier lowering (DIBL) effect, which is the most dominant characteristic of nanoscale devices in the near/sub-threshold region. Moreover, our $I_{ds}$ model achieves a great balance between precision and simplicity. It can be used for hand-calculation and early-stage design analysis.
- The non-Gaussian distributed circuit metrics in SRAM, such as bit-line (BL) differential voltage and write delay, are derived analytically and expressed by Gaussian random variables based on the drain current model. The corresponding distribution parameters in their probability density functions (PDF) can be characterized by trivial simulation efforts.
- Finally, the analytical models are constructed to evaluate probabilities of both the access-time and the write failures for SRAM cells with a remarkable speedup. Due to their good scalabilities, the proposed models can be directly used by SRAM comprised of different read/write assist circuits. Moreover, they can also be extended easily for different types of cell topologies. By taking the impact of supply voltage, transistor size, load capacitance into account, the results provide physical insights and guidelines for both cell and peripheral circuit designs.

The remaining parts of this paper are arranged as follows: Section II introduces the preliminaries, Section III develops the compact drain current model, access-time and write failure models sequentially, Section IV shows evaluation results, and Section V concludes the paper.

## II. Preliminaries

### A. *Inversion-charge based $I_{ds}$ models*

The classical textbook $I_{ds}$ model [16] for MOS transistors is widely used by other analytical models for logic path and SRAM [7]:

$$I_{ds} = I_0 e^{\frac{V_{gs}-V_{th}}{n v_t}} \cdot e^{\frac{\lambda V_{ds}}{n v_t}} \left(1 - e^{-\frac{V_{ds}}{v_t}}\right) \tag{1}$$

where $V_{th}$ is the threshold voltage, $n$ is the subthreshold swing parameter, $\lambda$ is the DIBL coefficient, and $v_t$ is the thermal voltage. Although (1) is simple and suitable for hand-calculation, it introduces strike errors in the near-threshold region due to the oversimplified relationship between $I_{ds}$ and $V_{gs}$.

A recent work [17] presents a new approximation of this relation via a Taylor series expansion, resulting in $I_{ds} \propto e^{K_1\left(\frac{V_{gs}-V_{th}}{n v_t}\right)+K_2\left(\frac{V_{gs}-V_{th}}{n v_t}\right)^2}$, where the $I_0$, $K_1$, $K_2$ are fitting constants. This approximation yields higher accuracy and continuity in the near-threshold region compared to the model (1) and is adopted by other works [10][11]. However, the impact of the lateral field gradient is neglected in [17]. Consequently, the short-channel effects, such as drain-induced barrier lowering (DIBL), are not accurately incorporated [18].

### B. *Failure Mechanisms of SRAM*

The random variations in process parameters during chip fabrication include inter-die (global) and intra-die (local) variations. The latter can be further divided into random variations and spatially correlated variations. The spatially correlated variations do not result in large differences between two transistors that are in close spatial proximity. While the random component of intra-die variations can result in a significant mismatch of threshold voltages between the neighboring transistors in a die. It will cause functional failures of cells in SRAM circuits. Here we briefly discuss the mechanisms of these failures.

#### 1) *Access Time Failure.*

Before accessing the cell, a pair of bit-lines are first precharged to $V_{DD}$ by the precharge PMOS. Then, the access transistors M2 and M3 are turned on while reading the cell shown in Fig. 1 ($V_Q$ = “1” and $V_{QB}$ = “0”). Transistor M3 will try to pull the bit-line BLB to the ground, while the voltage at BL remains high. After some time, a sense amplifier (SA) will detect the differential between BL and BLB. However, SAs also suffer from the $V_{th}$ mismatch that makes them have a "bias" (offset) voltage. The offset voltage can also be positive or negative. Only the input differential larger than the offset voltage can be detected as “1”, otherwise as “0”. Within a tolerable WL assertion time ($T_{WL}$), as illustrated in Fig. 1, the access time failure occurs when the difference between BL and BLB is below the certain positive offset voltage of the activated SA. This failure is caused by the strength reduction of the access and the pull-down NMOS (M3 and M1, or M2 and M0).

#### 2) *Read Failure.*

In Fig. 1, due to the voltage divider action between M1 and M3, the voltage at node QB ($V_{QB}$) increases to a positive value $V_{READ}$. If $V_{READ}$ is higher than the trip-point of the inverter M4-M0 ($V_{TRIP_L}$), then the very act of reading unintentionally over-writes the cell. This represents a read-failure event. A relatively stronger access NMOS (M3 or M2) and a weaker pull-down NMOS (M1 or M0) increase $V_{READ}$, thereby leading to a probability increase of the read failure.

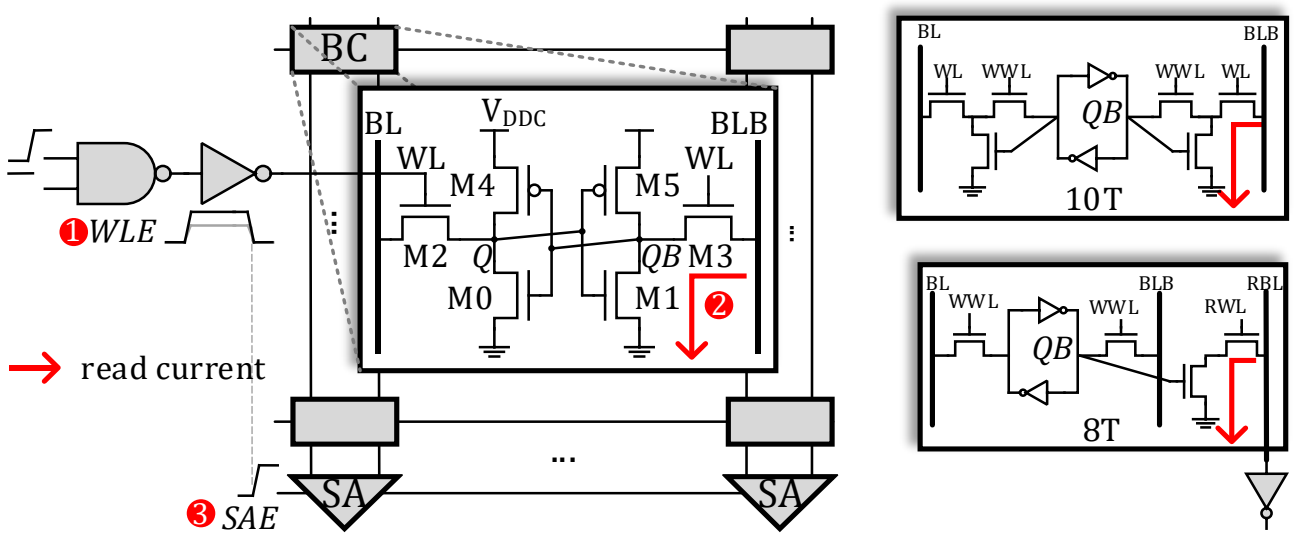

Fig. 1. Schematic of SRAM in a read operation. Assume the cell stores ‘1’ and $V_{QB}$ is low.

### 3) *Write Failure.*

While writing a “0” to a cell storing “1,” the node Q gets discharged through BL to a low value determined by the voltage division between the pull-up PMOS M4 and the access NMOS M2. At the same time, the node QB also is charged by BLB from “0” to the voltage divided by M3 and M1. If $V_Q$ cannot be reduced below $V_{TRIP_R}$ as well as $V_{QB}$ cannot be pumped above $V_{TRIP_L}$ within the time when word-line is high ($T_{WL}$), then a write failure occurs (Fig. 2). At the node QB, $V_{QB} > 0$ due to the voltage division, therefore, M3 always has a forward body bias, making its $V_{th}$ increase during the entire write operation (the body connects to the ground in SRAM cell). This results in the write current through M3 being smaller than that through M2. Therefore, discharging the node Q is faster than charging the other side, which means $V_Q$ is more likely to reach $V_{TRIP_R}$ and triggers the positive feedback of the SRAM cell (successful write).

Hence, a stronger PMOS and a weaker access transistor can significantly slow down the discharging process, thereby causing a write failure.

### 4) *Hold Failure.*

In the standby mode, the supply voltage of SRAM is reduced to minimize the leakage power consumption. However, if the lowering of $V_{DD}$ causes the data stored in the cell to be destroyed, then it is said to have failed in the hold mode. As the supply voltage of the cell is lowered, the voltage at the node storing “1” (node Q in Fig. 1) also gets reduced. Moreover, leakage of the pull-down NMOS M0 also reduces $V_Q$. If the voltage is reduced below $V_{TRIP_R}$, then flipping occurs and the data is lost in the hold mode.

Although all these failures may occur due to the process parameter variations, our work focuses on the most dominant failure at near-threshold voltages that relates to the $V_{DD}$ and temperature variations. Therefore, we set $1.3 \times 10^7$ MC simulations to compare the probabilities and the sensitivities to the PVT conditions of all those 4 types of failures. TSMC 28nm technology is used and the SRAM demo is provided in the process design kit (PDK). The access-time constraint is set to 4ns and the write-time constraint is set to 0.2ns. Table I shows

TABLE I

(A) PROBABILITY COMPARISON OF DIFFERENT TYPES OF FAILURES ACROSS DIFFERENT VOLTAGES.

| $V_{DD}$ (V) @ 25°C | 0.2 | 0.3 | 0.4 | 0.5 | 0.6 |
|---|---|---|---|---|---|
| Hold failures | 0 | 0 | 0 | 0 | 0 |
| Access-time failures | 1 | 1 | 0.154 | 1.37E-5 | 0 |
| Read failures | 0.028 | 2.78E-4 | 6.60E-6 | 0 | 0 |
| Write failures | 1 | 1 | 0.716 | 5.25E-4 | 0 |

(B) PROBABILITY COMPARISON OF DIFFERENT TYPES OF FAILURES ACROSS DIFFERENT TEMPERATURES.

| Temp. (°C) @ 0.5V $V_{DD}$ | -75 | -25 | 25 | 75 | 125 |
|---|---|---|---|---|---|
| Hold failures | 0 | 0 | 0 | 0 | 0 |
| Access-time failures | 0.116 | 3.90E-3 | 1.37E-5 | 0 | 0 |
| Read failures | 1.86E-4 | 4.01E-5 | 0 | 0 | 0 |
| Write failures | 1 | 0.84 | 5.25E-4 | 0 | 0 |

the probability comparisons of different types of failures (the number of failures divided by $1.3\times10^7$). Firstly, we find no hold failure occurs during our simulations with $V_{DD}$ rising from 0.2V to 0.6V and temperature increasing from -75°C to 125°C. It benefits from the low leakage current provided by the advanced nanometer technology. Secondly, the probability of read failure (read upset) is several orders of magnitude lower than that of the access failures. It increases with the lowered operating voltage and temperature. Thirdly, write failures and access-time failures are very sensitive to the changes of the operating voltage and temperature. Their probabilities increase abruptly when the $V_{DD}$ scales down from 0.5V to 0.4V. For the write failure, the probability dramatically rises as the temperature drops below 25°C. Lastly, compared to the access time, the minimum write time is much shorter at the same cell yield requirement.

To analysis these two SRAM failures, designers cannot only rely on the read/write static noise margins (SNMs) because they are used with the assumption that timing events have an infinite time duration ($T_{WL} = +\infty$) [8]. Dynamic metrics, such as access-time and write failures, represent the read and write behaviors in state-of-the-art SRAM designs with read-write-assist circuitry [19] and/or shrinking access cycle time [20][21]. Moreover, the accurate estimation of SRAM dynamic stability enables the evaluation and comparison of different dynamic circuit techniques, achieving agreement between estimates and measurements, fewer design iterations, and less time-to-market [22]. Thus, we construct analytical models to describe and abstract the mechanism of access-time and write failures in this work.

## C. *Modeling Access-time Failure*

Due to the high speed and storage density, 6T SRAM is widely used as the high-capacity on-chip storage [19] and can also be operated at near-threshold voltages [20][21][23] with simple assist circuits. For more complicated cell structures with separated read ports, such as single-port 8T [24] and 10T [25] illustrated in Fig. 1 and 2, their read paths comprised of an access transistor and a pull-down transistor are identical to that of 6T. The only difference is that the single-read-port 8T cell [24] employs an inverter or other logical gate to detect the voltage drop on the read bitline. For simplicity, we use a standard 6T SRAM cell as an example in the rest of the paper. However, a similar modeling process can be reproduced in other complicated scenarios and left for our future work.

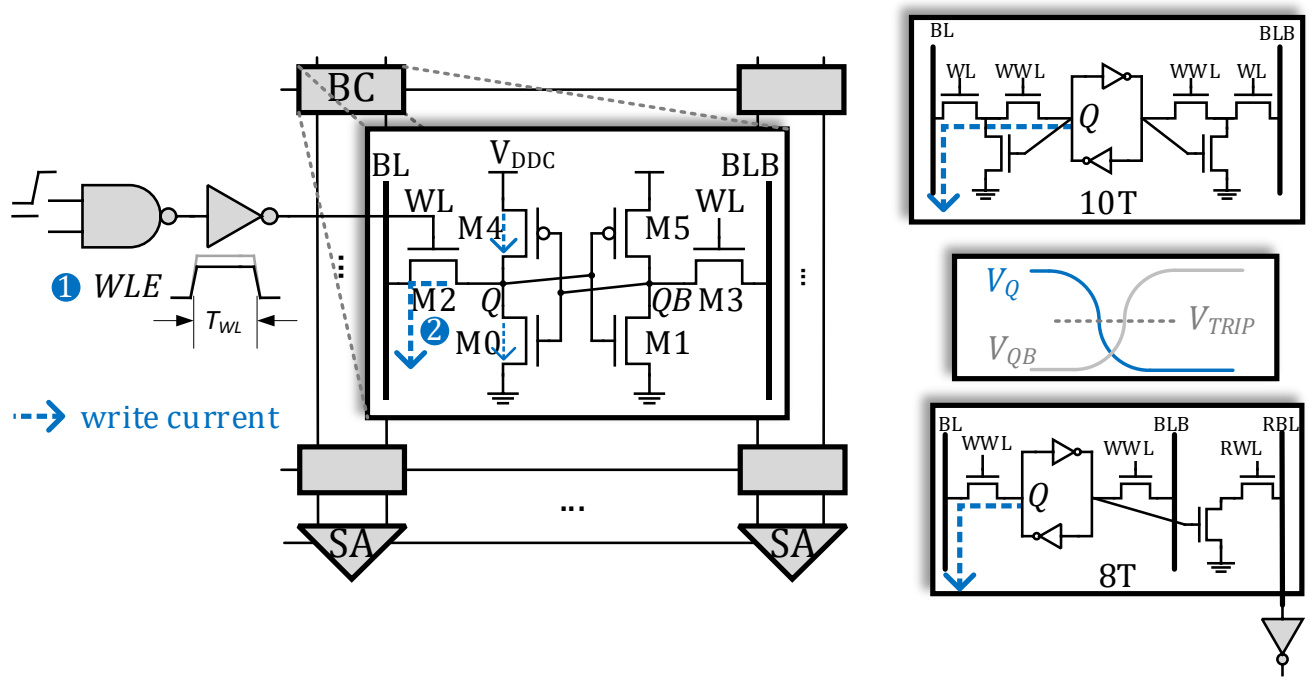

Fig. 2. Schematic of SRAM in a write operation. Assume we write a '0' into a '1' cell.

Complete SRAM access can be divided into several cycles to satisfy the core frequency. However, the critical path always contains the following steps as illustrated in Fig. 1 (assume the cell stores a '1'): ❶ the WL is first driven to high, ❷ the BLB with a large load starts to discharge from $V_{DD}$ to the ground through M3 and M1, and ❸ SAE activates the SAs [3]. At near-threshold voltages, discharging the bit-lines with large capacitances through those minimum-sized transistors becomes slower due to the reduced $I_{READ}$. The status of M2 is unchanged (its Vd and Vs are high) so that the voltage difference is mainly contributed by discharging BLB. As we discussed in Section II.B.1), by setting $T_{WL}$ to a specified time constraint ($T_{READ}$), then, the probability of an access-time failure emergence can be calculated as [1]

$$P_f = P(\Delta v < v_{OS} \mid T_{READ})$$
$$P_f = \Phi_\Delta(v_{OS}, T_{READ}) = \begin{cases} \int_{-\infty}^{v_{OS}} f_\Delta(\Delta v, T_{READ}) d\Delta v\,, & v_{OS} > 0 \\ 0, & v_{OS} \le 0 \end{cases} \quad (2)$$

where $\Delta v$ is differential voltage, $v_{OS}$ is the specified offset voltage of the SA, $f_\Delta$ , and $\Phi_\Delta$ are the PDF and the cumulated density function (CDF) of $\Delta v$. Prior works [1][2][3] treat $\Delta v$ (or $I_{READ}$) as a Gaussian random variable according to simulation results at the nominal voltage. However, this assumption no longer holds in sub/near-threshold region, leading to an increased error as supply voltage scales down [5]. Each cell failure event is related to those of other cells that share the same SA. This relationship among cell failures can be described by a joint PDF and the independence of $v_{OS}$ and $\Delta v$. The offset voltage is mainly caused by $V_{th}$ mismatch from the cross-coupled inverters in the SA, and it also follows a normal distribution even under low voltages [3]. Some classical models [1][2] assume $v_{OS}$ to be the worst-case, leading to a pessimistic yield estimation. Finally, the final fault probability for a random cell in an SRAM chip, or called bit error rate (BER), can be expressed by

$$P_f(T_{READ}) = \int_{-\infty}^{+\infty} f_{V_{OS}}(v_{OS}) \cdot \Phi_\Delta(v_{OS}, T_{READ}) dv_{OS} \quad (3)$$
$$f_{V_{OS}}(v) = \frac{1}{\sqrt{2\pi}\sigma_{V_{OS}}} e^{-\frac{(v-\mu_{V_{OS}})^2}{2\sigma_{V_{OS}}^2}}$$

where $f_{V_{OS}}$ is the PDF, $\mu_{V_{OS}}$ and $\sigma_{V_{OS}}$ are the mean and standard deviation of $v_{OS}$.

### D. Modeling Write-time Failure

Fig. 2 shows writing a "0" to a cell storing "1", where ❶ the wordline (WL) is asserted, and ❷ the node Q gets discharged while node QB gets charged through access transistors and bit-lines (10T [25] has 2 access NMOS on the write path). As we discussed in Section II.B.2), if the voltage of the discharged side cannot be below the $V_{TRIP}$ of the other-side inverter (M5-M1) within a given $T_{WL}$, a write failure occurs definitely. Thus, the minimum write time required by a write operation is [1][2]

$$t = \left| C_Q \int_{V_{DD}}^{V_{TRIP}} \frac{dV_Q}{I_{M4} - I_{M2}} \right| \quad (4)$$

where $I_{M4}$ and $I_{M2}$ are the channel current for M4 and M2 respectively. The current of M0 can be ignored compared to $I_{M4}$ and $I_{M2}$. $C_Q$ is the load capacitance of node Q. Given a specific time constraint ($T_{WL} = T_{WRITE}$), the probability of write failure can be calculated as [1]:

$$P_f(T_{WRITE}) = \int_{T_{WRITE}}^{\infty} f_W(t) dt \quad (5)$$

where $f_W$ is the PDF of write time.

Failure model [1] for SRAM yield analysis assumes $T_{WRITE}$ follows non-central *F* distribution, while other models for STA find the delay time of inverter chain follows Log-Normal [10], or the Log-Skew-Normal distribution [11]. Rather than caring about the specific shape of the PDF of access or write time, the importance-sampling-based methods have been proposed to construct a distorted sampling distribution by shifting the mean-vector (a vector containing process parameters of the 6 transistors in a cell) of the original distribution to the boundary or center of the failure region. The meta-model-based methods treat the relation between performance metrics and process parameters as a black-box, fitting the metrics with the non-linear-functions.

In general, the key for yield modeling is to find the exact PDF of the circuit metrics of interest, which is the main purpose of this work.

## III. Proposed Analytical Timing Yield Model

### A. $I_{DS}$ Model in Sub/near-threshold

In an NMOS transistor, the drain and source are connected by back-to-back *pn*-junctions (substrate-source and substrate-drain). In the vicinity of the *pn*-boundary, the electrons have been diffused away, or have been forced away by an electric field. The only elements left in this region are ionized donor or acceptor impurities. This region is called the depletion region.

Depletion regions of the source and reverse-biased drain junction become relatively more important with shrinking channel lengths (e.g., SRAM cell circuits). Since a part of the region below the gate is already depleted (by the source and drain fields), a smaller threshold voltage suffices to turn on the transistor. In other words, $V_{th0}$ decreases with channel length for short-channel devices. A similar effect can be obtained by raising the drain-source (bulk) voltage, as this increases the width of the drain-junction depletion region. Consequently, the threshold decreases with increasing $V_{ds}$, known as the effect of drain-induced barrier lowering (DIBL). It is the major mechanism in the near-threshold region that affects the output resistance in the saturation range [26][27].

Since the DIBL effect varies with the operating voltage, it is a problem in memories, where the read current of a cell (i.e., the drain current of the access transistor M3 in Fig. 1) becomes a function of the voltage on the bit-line, which complicates the differential voltage ($\Delta v$) modeling. Moreover, the leakage current from stand-by cells also depends upon the bit-line voltage and data patterns due to the DIBL. From the cell perspective, DIBL manifests itself as a data-dependent noise source [26].

Fortunately, we find that this effect can be modeled very well by introducing an extra exponential $V_{ds}$ term with the DIBL coefficient ($\lambda$). Another $V_{ds}$ term in our model that accounts for the transition between the linear and saturation regions is simplified and regulated by a constant $K_1$. Moreover, based on the work [17], the quadratic polynomial (involving $K_1$ and $K_2$) describing $I_{ds}$ changes with $V_{gs}$ is also adopted by our model. The final expression of drain current is

$$I_{ds} = I_0 e^{K_1(\frac{V_{gs}-V_{th}}{nv_t})+K_2(\frac{V_{gs}-V_{th}}{nv_t})^2} \cdot e^{\frac{\lambda V_{ds}}{nv_t}} \left(1 - e^{-K_1\frac{V_{ds}}{v_t}}\right) \quad (6)$$
$$I_0 = \mu \frac{W}{L} v_t^2 K_0$$

where $K_0$, $K_1$, $K_2$, and $\lambda$ are independent fitting constants, $\mu$ is the effective electron mobility, $n$ is the subthreshold swing parameter, and $v_t$ is the thermal voltage. Although this compact model introduces artificial constants besides the physical parameters, it shows a great approximation to the simulation results that will be discussed in Section IV.A.

### B. *Access-time Failure Model*

Bitline discharging process in SRAM access can be described as

$$-C_{BLB}\frac{dV_{BLB}}{dt} = I_{M3}(V_{gs} = V_{WL} - V_{QB},$$
$$-C_{BLB}\frac{dV_{BLB}}{dt} = I_{M3}(V_{ds} = V_{BLB} - V_{QB}, V_{bs} = -V_{QB}) \quad (7)$$

where $I_{M3}$ is the channel current of M3, $V_{WL}$ is the WL-driven voltage, $V_{QB}$ is the voltage of node QB in Fig.1. In this equation, the voltage drop on BL caused by cell leakage current is ignored in the target near-threshold region (thanks to the leakage control by the advanced technology). Moreover, given that the pull-down transistor M1 is stronger than M3 in a balanced design leading to a small $V_{QB}$, we ignore the substrate bias of M3. Then, (7) can be re-written by substituting (6) for $I_{M3}$:

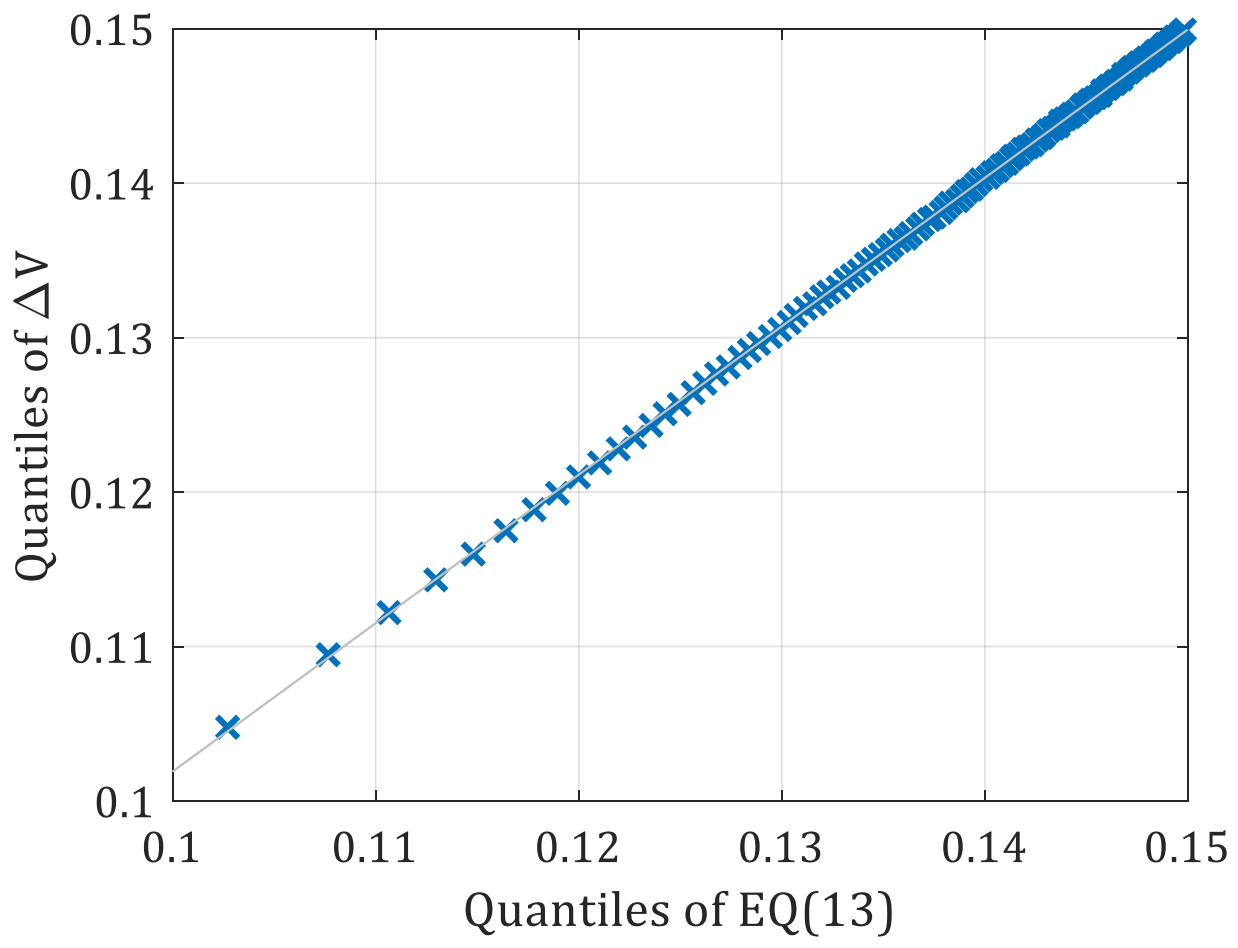

Fig. 3. Q-Q plot of $\Delta V$ at 0.5V 25°C TTG, with $T$ = 5.1ns.

$$-C_{BLB}\frac{d(V_{DD}-\Delta v)}{dt} = C_{BLB}\frac{d\Delta v}{dt} \quad (8)$$
$$= I_0 e^{K_1(\frac{V_{WL}-V_{th}}{nv_t})+K_2(\frac{V_{WL}-V_{th}}{nv_t})^2} \cdot e^{\frac{\lambda(V_{DD}-\Delta v)}{nv_t}} \cdot \left(1 - e^{-K_1\frac{V_{DD}-\Delta v}{v_t}}\right)$$

where $\Delta v$ is the BL-BLB differential voltage, $C_{BLB}$ is the equivalent load capacitance of BLB. The last factor $\left(1 - e^{-K_1\frac{V_{DD}-\Delta v}{v_t}}\right)$ is very close to 1 because of the small $\Delta v$ (usually tens of millivolt). Let $p(V_{gs}, V_{th}) = K_1\left(\frac{V_{gs}-V_{th}}{nv_t}\right) + K_2\left(\frac{V_{gs}-V_{th}}{nv_t}\right)^2$ and integrate both sides of (8):

$$C_{BLB}\int_0^{\Delta V} e^{\frac{\lambda(\Delta v - V_{DD})}{nv_t}} d\Delta v = I_0 e^{p(V_{WL},V_{thn})} \cdot \int_0^{T_{READ}} dt \quad (9)$$
$$\Rightarrow C_{BLB}\frac{nv_t}{\lambda}\left(e^{\frac{\lambda(\Delta V - V_{DD})}{nv_t}} - e^{-\frac{\lambda V_{DD}}{nv_t}}\right) = I_0 e^{p(V_{WL},V_{thn})} \cdot T_{READ}$$

where $T_{READ}$ is the pre-set time constraint. After calculating the logarithm of (9) and bringing $\Delta V$ to the left side, we get

$$\Delta V = \frac{nv_t}{\lambda_n} ln\left(e^p(\alpha T_{READ} + e^{-\frac{\lambda V_{DD}}{nv_t}-p})\right) + V_{DD} \quad (10)$$

where $\alpha = \frac{\lambda I_0}{nv_t C_{BLB}}$. Compared to the variation of $p(V_{WL}, V_{thn})$ in (10), $-\frac{\lambda V_{DD}}{nv_t}$ dominants the exponential term while $\alpha\ T_{READ}$ dominates the total value in the parenthesis. Therefore, we can approximate the second $p$ in (10) to its mean value $p_0$ yielding

$$\Delta V \approx \frac{nv_t}{\lambda} p(V_{WL}, V_{thn}) + g(T_{READ}) \quad (11)$$
$$g(t) = \frac{nv_t}{\lambda} ln\left(\alpha t + e^{-\frac{\lambda V_{DD}}{nv_t}-p_0}\right) + V_{DD}$$

where $g(T_{READ})$ is a function of the specified access time and other process parameters. From (11), it's obvious that

$$\Delta V \propto p(V_{WL}, V_{thn}) \Rightarrow \Delta V \propto K_2 \left(\frac{V_{WL}-V_{thn}}{nv_t} + \frac{K_1}{2K_2}\right)^2 \quad (12)$$

Here, we only consider the $V_{th}$ variations in our model because, firstly, the threshold variation is the most dominant factor to cause a cell hard failure (the same conclusion drawn from [1]-[3][10][11][13][14]). Secondly, introducing several random variables in our model will complicate the expression of the metrics of interest, and result in difficulty to find their PDFs. In MOSFET models [26][27], the threshold voltage $V_{thn}$ is a linear function of the threshold voltage for zero substrate bias ($V_{th0}$). $V_{th0}$ is one of the Gaussian-distributed process parameters in the technology library provided by the foundry. Consequently, $V_{thn}$ is a Gaussian random variable. By treating $V_{WL}$ as a fixed value ($V_{DD}$), $\Delta V$ becomes a special case of a non-central Chi-square random variable with 1 degree of freedom. The non-Gaussian $\Delta V$ can be expressed by a Gaussian one, $\sqrt{\Delta V} \sim N(\mu_\Delta, {\sigma_\Delta}^2)$, which matches the observation reported by [5]. Through a simple manipulation, the PDF of $\Delta V$ is

$$f_\Delta(\Delta v) = \frac{1}{2\sigma_\Delta\sqrt{2\pi\Delta v}} e^{-\frac{(\sqrt{\Delta v}-\mu_\Delta)^2}{2\sigma_\Delta^2}} \quad (13)$$

where parameters $\mu_\Delta$ and $\sigma_\Delta$ are the mean and standard deviation of $\sqrt{\Delta V}$. They can be directly collected in Monte Carlo simulations at different transient times in SPICE. By taking (13) into (3) and (4), the final yield at a given $T_{READ}$ can be obtained.

Fig. 3 shows the Q-Q plot of $\Delta V$ at TTG. We use $1.3\times10^7$ $\Delta V$ samples from MC simulations at 0.5V with $T_{WL}$ = 5.1ns to quantify $\mu_\Delta$, $\sigma_\Delta$, and the quantiles at the y-axis. It shows the tail of quantiles, where the small $\Delta V$ accounts for the access failures. The converging line of quantiles indicates the distribution of data samples is linearly related to the modeled one, which means (13) can depict the distribution of $\Delta V$ accurately.

### C. *Write Failure Model*

To evaluate write-time failures, the current in (4) needs to be replaced by our drain current model. The transregional term $\left(1 - e^{-K_1\frac{V_{ds}}{v_t}}\right)$ is close to 1 and can be ignored as long as $V_{ds} > 3v_t$ (thermal voltage ≈ 25mV). During discharging node Q from $V_{DD}$ to $V_{TRIP}$, $V_{TRIP}$ varies in the range of (40% $V_{DD}$, 62% $V_{DD}$) after our evaluation (the target $V_{DD}$ is from 0.5V to 0.7V). So we remove the transregional term in the $I_{ds}$ equations. By using subscripts of the process parameters to distinguish PMOS M4 and NMOS M2, current at node Q can be expressed as:

$$-I_{M2} + I_{M4} = -I_{0n} \cdot e^{p_n} \cdot e^{\frac{\lambda_n V_Q}{n_n v_t}} + I_{0p} \cdot e^{p_p} \cdot e^{\frac{\lambda_p (V_Q - V_{DD})}{n_p v_t}} \quad (14)$$
$$-I_{M2} + I_{M4} = -e^{p_n}\left(I_{0n} \cdot e^{\frac{\lambda_n V_Q}{n_n v_t}} - \beta I_{0p} \cdot e^{\frac{\lambda_p (V_Q - V_{DD})}{n_p v_t}}\right)$$

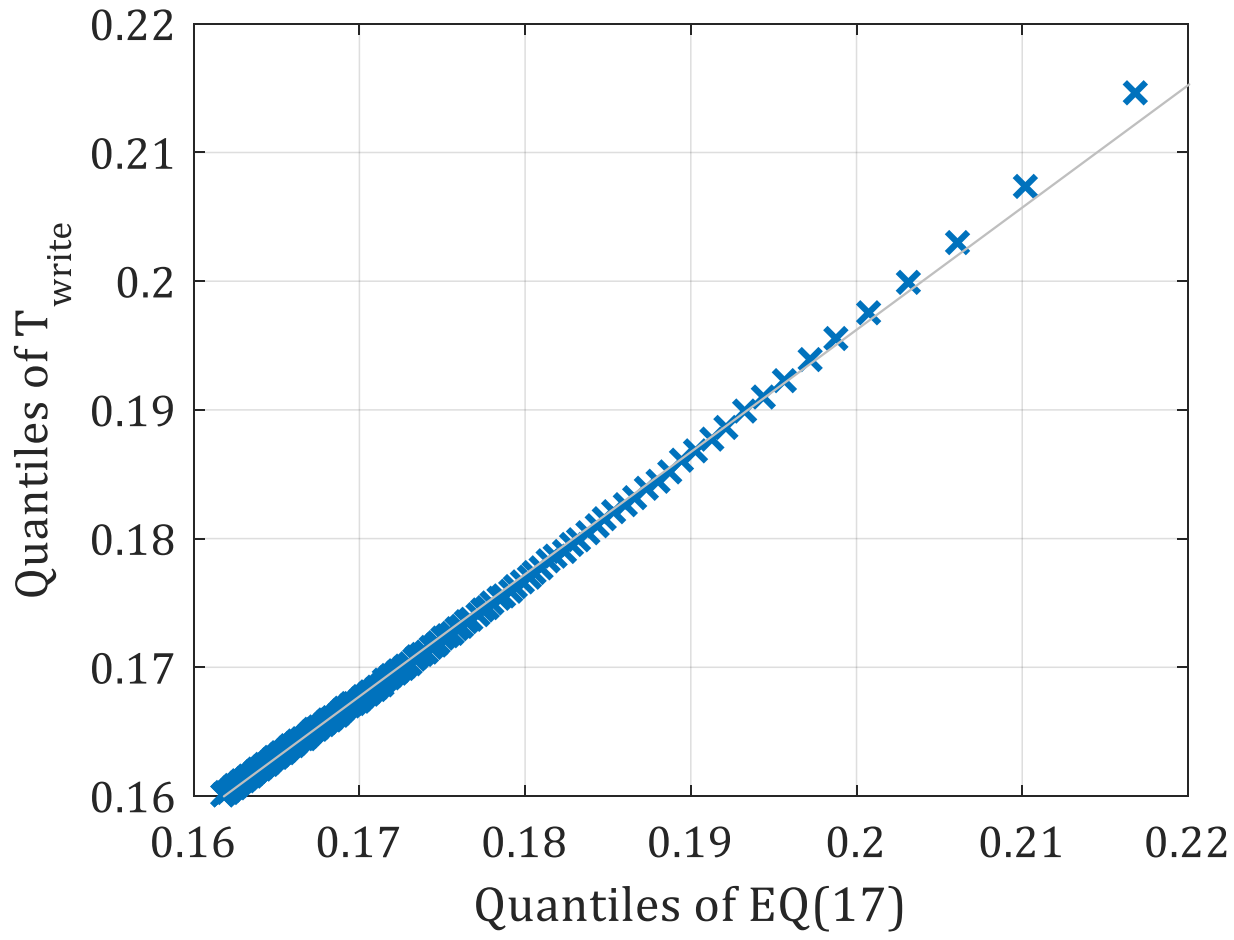

Fig. 4. Q-Q plot of $T_{WRITE}$, at 0.5V TTG 25°C.

where $p_n(V_{WL}, V_{thn}) = K_{1n}(\frac{V_{WL}-V_{thn}}{n_n v_t}) + K_{2n}(\frac{V_{WL}-V_{thn}}{n_n v_t})^2$, $p_p(V_{QB} - V_{DD}, V_{thp}) = K_{1p}(\frac{V_{QB}-V_{DD}-V_{thp}}{n_p v_t}) + K_{2p}(\frac{V_{QB}-V_{DD}-V_{thp}}{n_p v_t})^2$, and $\beta = e^{p_p - p_n}$. Because the first term is several times larger than the second in the parenthesis, same as the approximation done in (11), we replace $\beta$ with its mean value $\beta_0$. Then, take (14) into (4) and separate $e^{p_n}$ from the denominator, the minimum write time becomes

$$t \approx -C_Q e^{-p_n(V_{WL}, V_{thn})} \cdot w(V_{TRIP}) \quad (15)$$
$$w(v) = \int_{V_{DD}}^{v} \frac{dV_Q}{I_{0n} \cdot e^{\frac{\lambda_n V_Q}{n_n v_t}} - \beta_0 I_{0p} \cdot e^{\frac{\lambda_p (V_Q - V_{DD})}{n_p v_t}}}$$

where the $w(V_{TRIP})$ is the function of the trip-point voltage of the right-side inverter depicted in Fig. 2 and other process parameters. Variations of transistor M3 and M1 in the right side of a cell are incorporated in $V_{TRIP}$, making it a random variable. However, similar to prior work [1][2], we neglect the variation of $V_{TRIP}$, because the $w(V_{TRIP})$ in (15) does not have an explicit form, which would be very hard to derive its PDF. Furthermore, modeling variations of M2 and M4 are more important than those of M3 and M1 because discharging the node Q is faster than charging the node QB and determines the minimum write time (as discussed in Section II.B). Fortunately, the simplification only results in a little underestimation (Fig. 10).

Now the logarithm of $t$ is proportional to $p_n(V_{WL}, V_{thn})$,

$$t \propto e^{-p_n(V_{WL}, V_{thn})} \quad (16)$$
$$\Rightarrow ln(t) \propto -K_{2n}\left(\frac{V_{WL}-V_{thn}}{n_n v_t} + \frac{K_{1n}}{2K_{2n}}\right)^2$$

Due to $V_{WL}$ being a fixed value and $V_{thn}$ being an independent Gaussian random variable, the logarithm of write time conforms to a non-central Chi-square distribution with 1 degree of freedom. The non-Gaussian $t$ can also be transformed from a Gaussian random variable, $\sqrt{ln(t)} \sim N(\mu_W, {\sigma_W}^2)$. The final

PDF of $T_{WRITE}$ equals to

$$f_W(t) = \frac{1}{2\sigma_W \cdot t\sqrt{2\pi\cdot \ln(t)}} e^{-\frac{(\sqrt{\ln(t)}-\mu_W)^2}{2\sigma_W^2}} \tag{17}$$

where the $\mu_W$ and $\sigma_W$ are the mean and standard deviation of $\sqrt{ln\,(t)}$. Finally, the write yield can be calculated through (17) and (5).

Fig. 4 shows the tail part of the whole Q-Q plot of $T_{WRITE}$. The parameters $\mu_W$ and $\sigma_W$ in (17) and the y-axis quantiles are calculated from 1.3 × $10^7$ samples of $T_{WRITE}$ using MC simulations at 0.5V 25°C TTG. Quantile samples approximately lie on a line demonstrating that (17) models the distribution of $T_{WRITE}$ accurately.

### D. Models with Read/Write Assist Circuits

Many SRAM designs use read-assist circuits to avoid read failures during access and enhance read stability. For instance, as Fig. 1 shows, WL under-drive [28] weakens the pass-gate transistor to keep $V_{QB}$ low. Cell $V_{DD}$ boost [29] increases the cell $V_{DD}$ with a separated power supply to strengthen the pull-down NMOS and increase $V_{TRIP}$ during reading. Our models can also be used in SRAM with read-assist techniques. When WL under-drive is employed, it is equivalent to reducing $V_{WL}$ by a fixed value. The resistance of M3 increases because of its lowered $V_{gs}$. Our assumption of the zero-substrate-bias of M3 still holds. Then, $V_{WL}$ in (8)-(11) is replaced with the reduced WL-driven voltage, $V_{WLUD}$. Equation (12) becomes:

$$\Delta V \propto K_2\left(\frac{V_{WLUD}-V_{thn}}{nv_t}+\frac{K_1}{2K_2}\right)^2 \tag{18}$$

Due to $V_{WLUD}$ being a fixed value, the shape of the PDF in (13) doesn't change. Similarly, when using separated cell supply voltage ($V_{DDC}$), the rise of $V_{DDC}$ only affects the value of $\mu_\Delta$ and $\sigma_\Delta$, while keeping our assumption and the form of PDF intact.

For write-assist techniques, as shown in Fig. 2, boosting the WL-driven voltage [30] helps with the write margin for enhancing the contention ability of the access NMOS M2 and speeding up the completion of writing a "1" to the other side. It can be considered as increasing $V_{WL}$ by a fixed value. The stronger M2 with a larger write current further reduces the error of our approximation in (15). By replacing $V_{WL}$ with $V_{WLB}$ in (14) and (15), the logarithm of minimum write time

$$l\,n(t) \propto -K_{2n}\left(\frac{V_{WLB}-V_{thn}}{n_n v_t}+\frac{K_{1n}}{2K_{2n}}\right)^2 \tag{19}$$

This doesn't change the form of PDF in (17). In another scenario, cell supply voltage collapse [31] suppresses the latching capability of an SRAM cell to make a fast and easy write operation. This method reduces the $V_{ds}$ and $V_{gs}$ of M4 to $V_Q$ - $V_{DDC}$ and $V_{QB}$ - $V_{DDC}$ respectively. The trip point of the right-side inverter is also reduced to $V_{TRIPC}$. Then, $w$ in (15) is re-written as

$$w(V_{TRIPC}) = \int_{V_{DD}}^{V_{TRIPC}} \frac{dV_Q}{I_{0n}\cdot e^{\frac{\lambda_n V_Q}{n_n v_t}} - \beta_0 I_{0p}\cdot e^{\frac{\lambda_p (V_Q - V_{DDC})}{n_p v_t}}} \tag{20}$$

However, all the small differences between the derived functions with or without write assist have no impact on our final conclusion but only the exact values of $\mu_W$ and $\sigma_W$ in (17). In the next section, we will validate our analytical yield models in the scenarios where the WL under-drive and boost techniques are applied in SRAM.

## IV. Evaluation

This section first shows the validation of our compact $I_{ds}$ model, then compares the proposed models from both accuracy and efficiency perspectives.

All experiments are performed with HSPICE on a server with Intel Xeon Gold 5118 CPU @ 2.30 GHz. We use the TSMC 28nm library based on the BSIM 4.5.0 model [27] to set up our experiments. There are total 402 process parameters for each transistor, including 78 global variables and 14 local variables, such as coefficient of the drain voltage reduction (*eta*), low field mobility at nominal temperature ($u_0$), electrical gate oxide thickness in meters ($t_{oxe}$), and channel doping concentration at the depletion edge for the zero body bias ($n_{dep}$). All process parameters' variabilities are turned on in our simulations. BSIM model [27] is an accurate device model and integrated into the SPICE tool. The process parameters in the technology library are provided by TSMC. Both global and local process parameters are Gaussian random variables whose mean values and standard deviations are also defined in the library. The variations of device geometry, spatial variations, and other correlated variations are also set in the technology library. The standard 6T SRAM demo is designed by TSMC existing in their PDK.

In the experiment part, we use different corners to prove that our cell failure models perform well under different global process variations. TTG, FFG, and FSG represent different

TABLE II

Errors of $I_{DS}$ Model (7) Used in PMOS (pch) and NMOS (nch) with High (hvt), Standard (svt), and Low Vth (lvt) Transistors at TTG 25°C.

| Param. | $I_0$ | $K_1$ | $K_2$ | $\lambda$ | Vgs=0.6V | | Vgs=0.4V | |
|---|---|---|---|---|---|---|---|---|
| | | | | | Max. error | Avg. error | Max. error | Avg. error |
| nch_hvt | 5.3796e-6 | 0.3981 | -0.0296 | 0.0201 | 7.7% | 3.7% | 6.6% | 3.3% |
| pch_hvt | 1.3225e-6 | 0.5068 | -0.0360 | 0.0315 | 12.0% | 3.4% | 10.8% | 3.2% |
| nch_svt | 2.29e-5 | 0.1414 | -0.0028 | -0.0012 | 5.8% | 3.9% | 9.1% | 4.0% |
| pch_svt | 7.9742e-6 | 0.3102 | -0.0093 | 0.0014 | 5.3% | 3.6% | 7.9% | 3.6% |
| nch_lvt | 1.4854e-5 | 0.3157 | -0.0118 | 0.0161 | 7.4% | 2.6% | 2.9% | 2.0% |
| pch_lvt | 1.5647e-5 | 0.2547 | -0.0082 | 0.0135 | 10.5% | 4.2% | 3.9% | 1.8% |

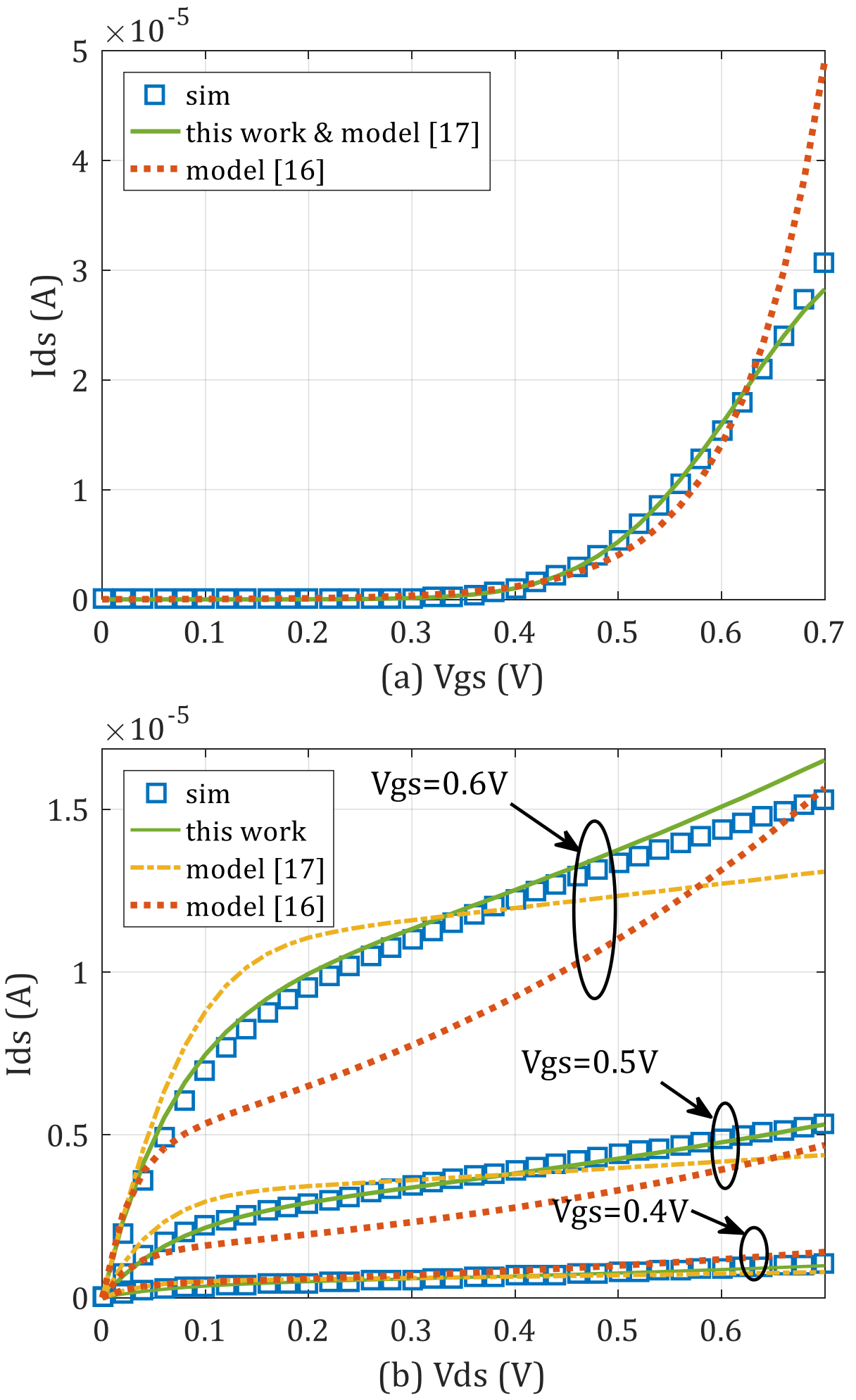

Fig. 5 (a) $I_{ds}$ vs. $V_{gs}$ with $V_{ds}$ = 0.7V and (b) $I_{ds}$ vs. $V_{ds}$ with $V_{gs}$ = 0.6V, 0.5V, 0.4V using 28nm technology, at 25ºC TTG corner.

types of global process corners, where TTG stands for typical NMOS typical PMOS, FFG for fast NMOS fast PMOS, and FSG for fast NMOS slow PMOS.

### A. Validation of the Compact $I_{ds}$ Model

We first collect $I_{ds}$ vs. $V_{gs}$ and $I_{ds}$ vs. $V_{ds}$ data through SPICE simulations with the DC analysis. The data set needs to be constructed for each type of device with a minimum size. The range of $V_{gs}$ and $V_{ds}$ changing is from 0V to 0.7V with a step of 0.01V. Then, the data set is input into MATLAB, the *nlinfit* procedure is applied to extract the process constants. The nominal supply voltage of the SRAM transistors is 0.8V.

Fitting constants in our compact model (6) are unrelated to variations of process parameters and their values are listed in Table II.

Fig. 5 (a) and (b) depict different compact drain-current models. Due to the oversimplified relation of inversion-charge and the terminal voltage, the classic model [16] introduces strike errors in the near-threshold region. The improved work [17] performs well in modeling $I_{ds}$ vs. $V_{gs}$, but not well in $I_{ds}$ vs. $V_{ds}$. This is because the DIBL effect makes the drain current a slope in the velocity saturated region. Our model shows a good match to simulation results across different $V_{gs}$ and $V_{ds}$.

Our compact model can also be used by other types of transistors with high, standard, and low $V_{th}$. Table II lists evaluation errors at TTG 25°C. The max/average relative errors are calculated from the saturated region (where $V_{dsat} < V_{ds}$). The max relative error is only 12% for the PMOS with a high threshold voltage (denoted as pch_hvt). The average relative error in the saturated region is always below 4.2% across different types of transistors. The decrease of average errors for low-$V_{th}$ devices at lower $V_{gs}$ is because (6) depicts the tendency of $I_{ds}$ more precisely at sub/near-threshold region (already shown in Fig. 5 (a)).

### B. Models Predictions vs. Monte Carlo Results

Simulation results are collected from a 256-row by 64-column sized array with a word-line driver. Criterions of all types of failures in our experiments have been discussed in Section II.B. All initial data of cells is set to "1" to match the assumption during modeling. For the access-time failure simulation, we first run MC simulations with sense amplifier circuits to obtain a data set of offset voltages. Then, the voltage differences $\Delta V$ between bit-lines are collected from an SRAM array and compared to the $V_{OS}$ in new MC simulations. For the write failure simulation, we conduct DC analysis for the right side of a cell, collecting samples of $V_{TRIP_R}$ through MC sweeps. Then, new MC simulations with transient analysis are set, and the voltages of the node Q of the written cell are compared to the corresponding values of $V_{TRIP_R}$ to find minimum write times.

Considering acceptable time/storage overhead, we run $1.3\times 10^7$ MC simulations that account for 4σ ($3.17\times10^{-5}$)[1] of failure estimation with a 95% confidence interval as the gold standard for each PVT. This magnitude of yield is also adopted by other yield analysis methods [5][7][12][13][14]. Note that our model has the capability to predict the failures at a higher-σ region, however, validations at 6σ ($9.87\times10^{-10}$) requires more than $10^{13}$ simulation samples that may take several years to run. The relative error is defined as:

$$error = \frac{|P_f - \widehat{P_f}|}{P_f} \times 100\% \quad (21)$$

where $P_f$ is the failure probability estimated by Monte Carlo simulations and $\widehat{P_f}$ is the prediction from our access-time or write failure models.

Fig. 6 (a) compares the relative errors at 4σ between other access-time failure models and ours. Failure model [1] has larger deviations across all operating voltages because it only considers the variations of read current ($I_{READ}$) and assumes access time as a Gaussian random variable. VAM [9] that only models $I_{READ}$ distribution also has large errors because $I_{READ}$ is a function of $\Delta V$ and $T_{WL}$. In work [3], the differential voltage $\Delta V$ is supposed to follow the Gaussian distribution, which results in the error growth at a lower $V_{DD}$. Our model improves the evaluation accuracy of [1], [3], and [9] by 31.6X, 17.6X, and 25.9X at 0.5V, respectively. Studies [12] and [13] mentioned in Section I are importance-sampling-based methods and achieve similar accuracy but with more data samples compared to our work (Table IV). Fig. 7 (a) shows failure probabilities down to $10^{-6}$ at 25°C TTG. Our work achieves great accuracy with 15.7% of maximum error and 8.8% of the average error in the wide range of $T_{READ}$ across different voltages. As the voltage reduces, SRAM access time satisfying 4σ yield is 7.85X at 0.5V compared to that at 0.7V $V_{DD}$. Such a performance degradation makes $V_{MIN}$ of SRAM module 100~200mV higher than $V_{MIN}$ of the logic module. Fig. 7 (b) further shows the yield changes with different temperatures at 0.5V FFG. The operating temperature also affects its performance and yield. The model prediction errors are 4.9%, 6.4%, 10.5%, and 12.7% in the range of ($10^{-5}$, $10^{-6}$) for -25°C, 25°C, 75°C, and 125°C respectively. At 75°C, the $T_{READ}$ is only half of that at 25°C, while increases to 1.41X at -25°C. This character is fully analyzed and utilized by several low-power SRAM designs [20][23].

[1] We use "4σ" here to represent the corresponding probabilities, $P(X \geqslant \mu+4\sigma)$, where $X$ is the data set of access (write) time, $\mu$ is the mean and $\sigma$ is the standard deviation of $X$. It doesn't necessarily mean Gaussian distributions.

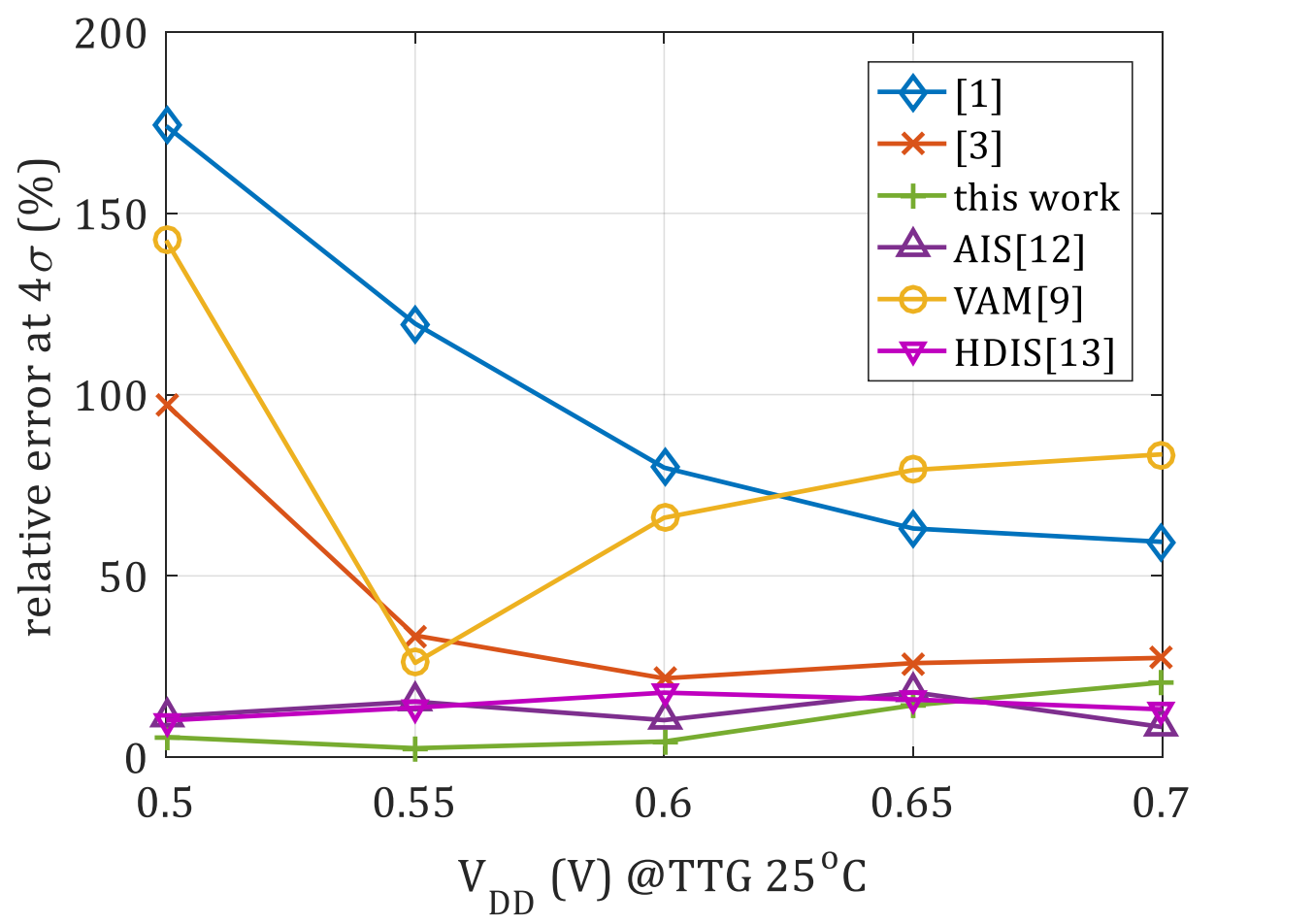

Fig. 6. Relative error comparison of analytical yield models at different $V_{DD}$ 25°C TTG corner.

Fig. 8 shows $T_{READ}$ at different WL voltages using WL under-drive assist circuits. The maximum relative error of our model is 9.7% at 0.525V $V_{WL}$. The read assist circuits sacrifice the access delay for reducing the read upset probability. A 100mV $V_{WL}$ reduction worsens $T_{READ}$ nearly 2X at 0.6V $V_{DD}$, by contrast, a 25mV reduction has no significant impact on the read speed. Such a tradeoff between WL voltage and access delay is the motivation of design [32].

Regarding write failure estimations, Fig. 9 compares different write failure models at 25°C TTG corner. Failure model [1] uses non-central *F* distribution to describe write time. Statistical timing models [10] and [11] for STA using the Log-Normal (LN) and Log-Skew-Normal (LSN) distributions yield similar precisions. Errors of their predictions grow slowly as $V_{DD}$ scales down, caused by the assumption of unchanged $V_{ds}$ and regular input pules. The accuracy improvements of our

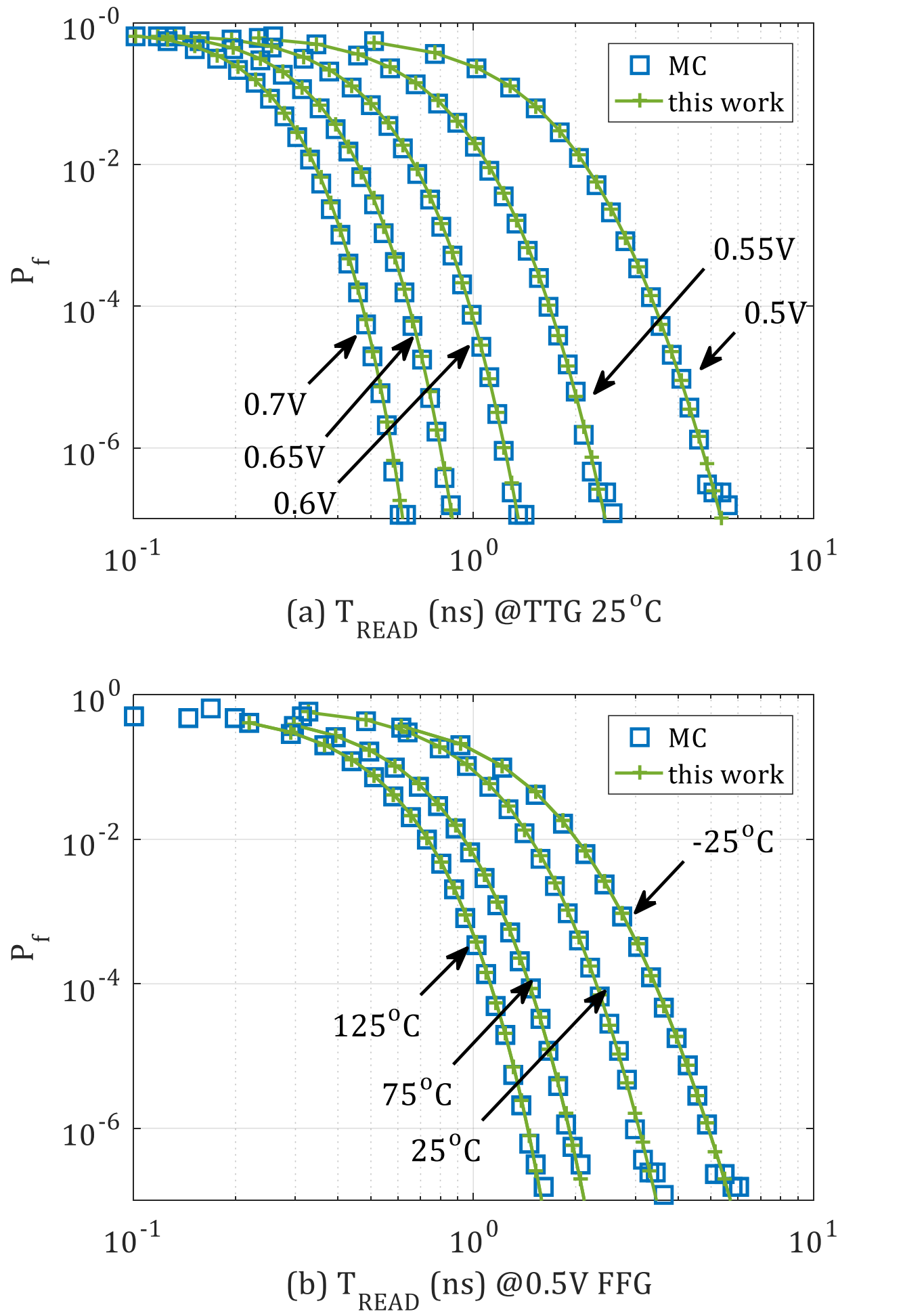

Fig. 7. Comparison between MC results and the proposed model in a wide range of timing constraints at (a) different $V_{DD}$ TTG corner and (b) different temperatures FFG corner.

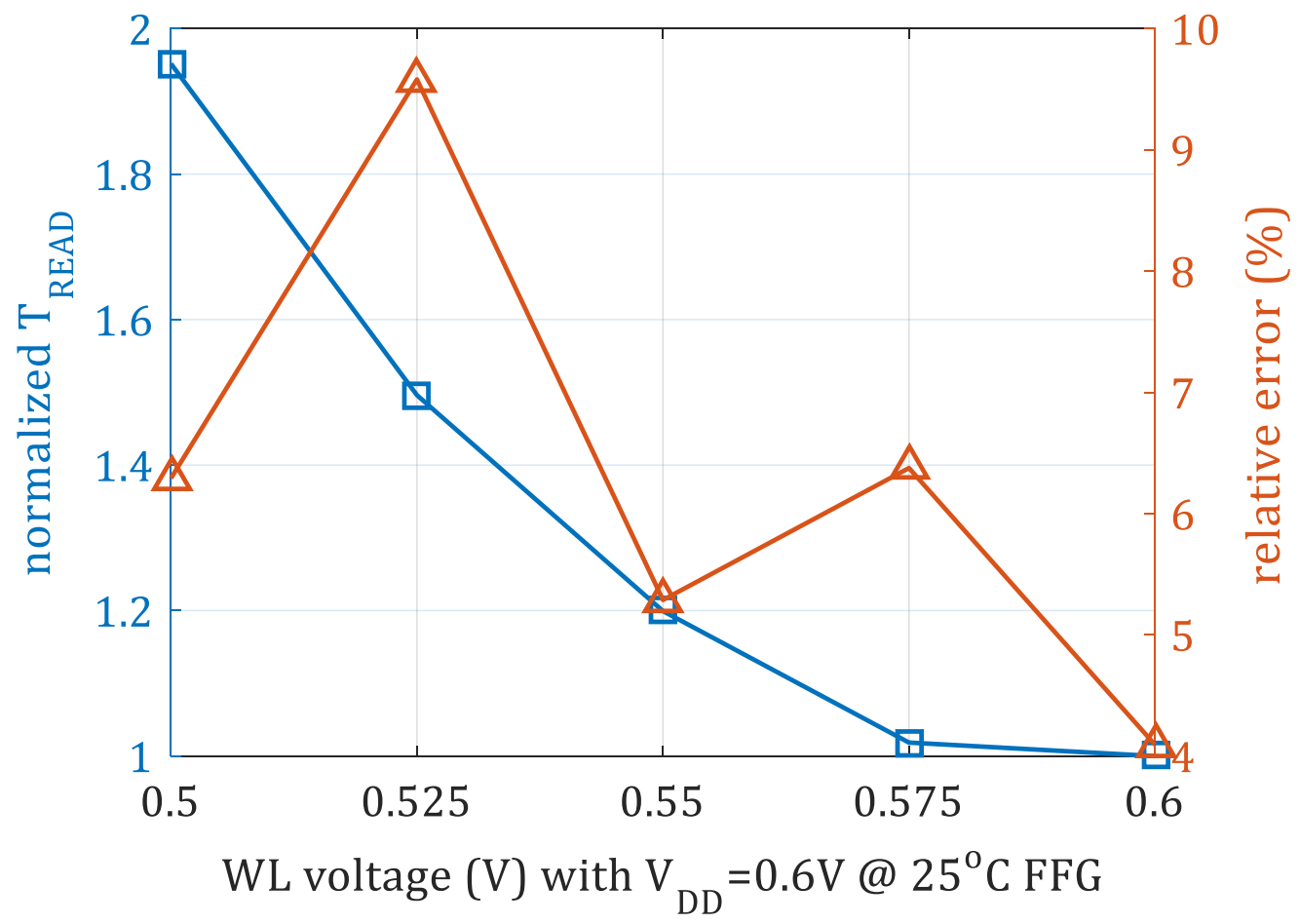

Fig. 8. Normalized $T_{READ}$ (left) and relative errors (right) of the proposed model at 25°C FFG corner.

model are 15X, 7.5X, and 6.3X at 0.5V compared to models [1], [10], and [11], respectively. AIS [12] shows its instability with 34.1% relative error at 0.7V in predicting the write-time failures. The maximum error of our model is 17.3% at 0.55V FFG. Fig. 10 (a) and (b) shows the model predictions and the MC simulation results at different $V_{DD}$ and temperatures. For a write

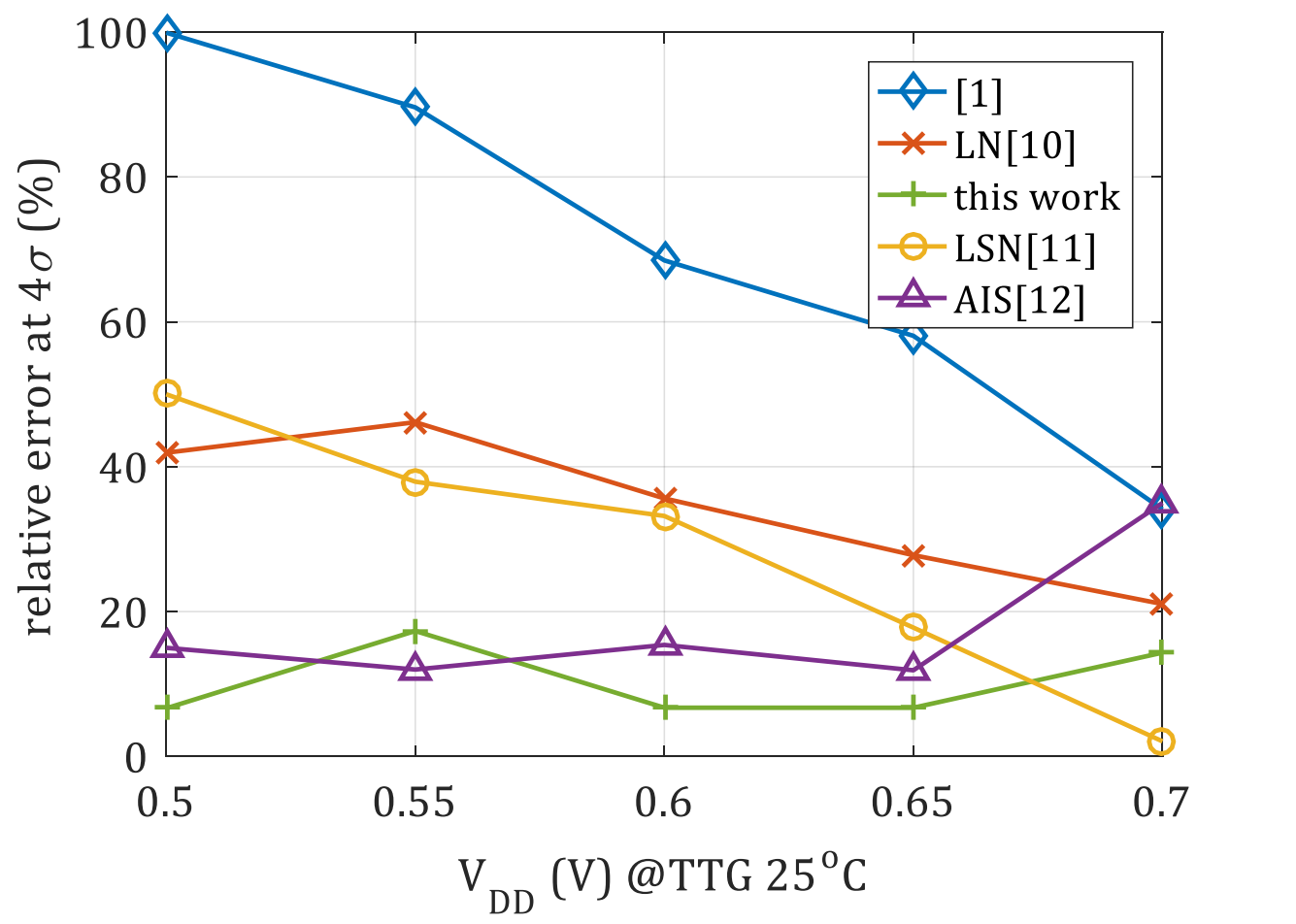

Fig. 9. Relative error comparison of analytical yield models at different $V_{DD}$ 25°C TTG corner.

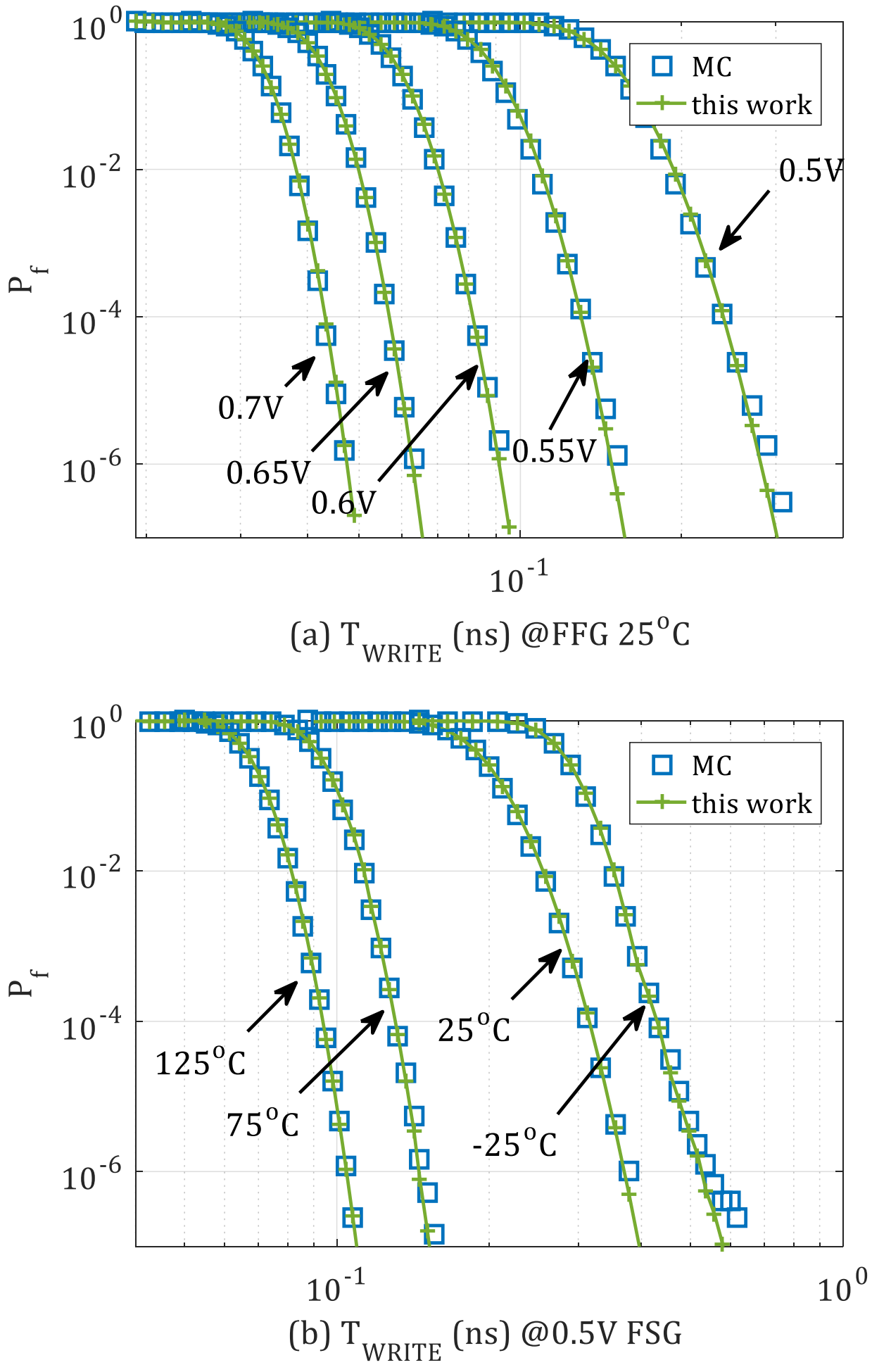

Fig. 10 (a) Comparison between MC results and the proposed model in a wide range of timing constraints at (a) different $V_{DD}$ FFG corner and (b) different temperatures FSG corner (with fast NMOS and slow PMOS).

operation, driving the internal storage node with small capacitance is much faster compared to SRAM read operation. Thus, $T_{WRITE}$ has a concentrated distribution with a very small mean and variance but a large skewness. Such a distribution increases the average errors of our model predictions and makes itself very sensitive to PVT conditions. For example, at 4σ region, the write delay is shortened to 0.52X when $V_{DD}$ scales from 0.5V to 0.55V and 0.42X when the temperature rises from 25°C to 75°C. However, the average error of our model across all PVT is always below 10.4% in a wide range of failure probabilities. The underestimation of our write failure model is mainly caused by ignoring the variations of $V_{TRIP_R}$ in (15), which is equivalent to ignoring those of the right side in a SRAM cell. Besides, the underestimated error grows as the voltage or temperature goes down. This demonstrates that a lower voltage or temperature enhances the impact of process variations on memory transistors.

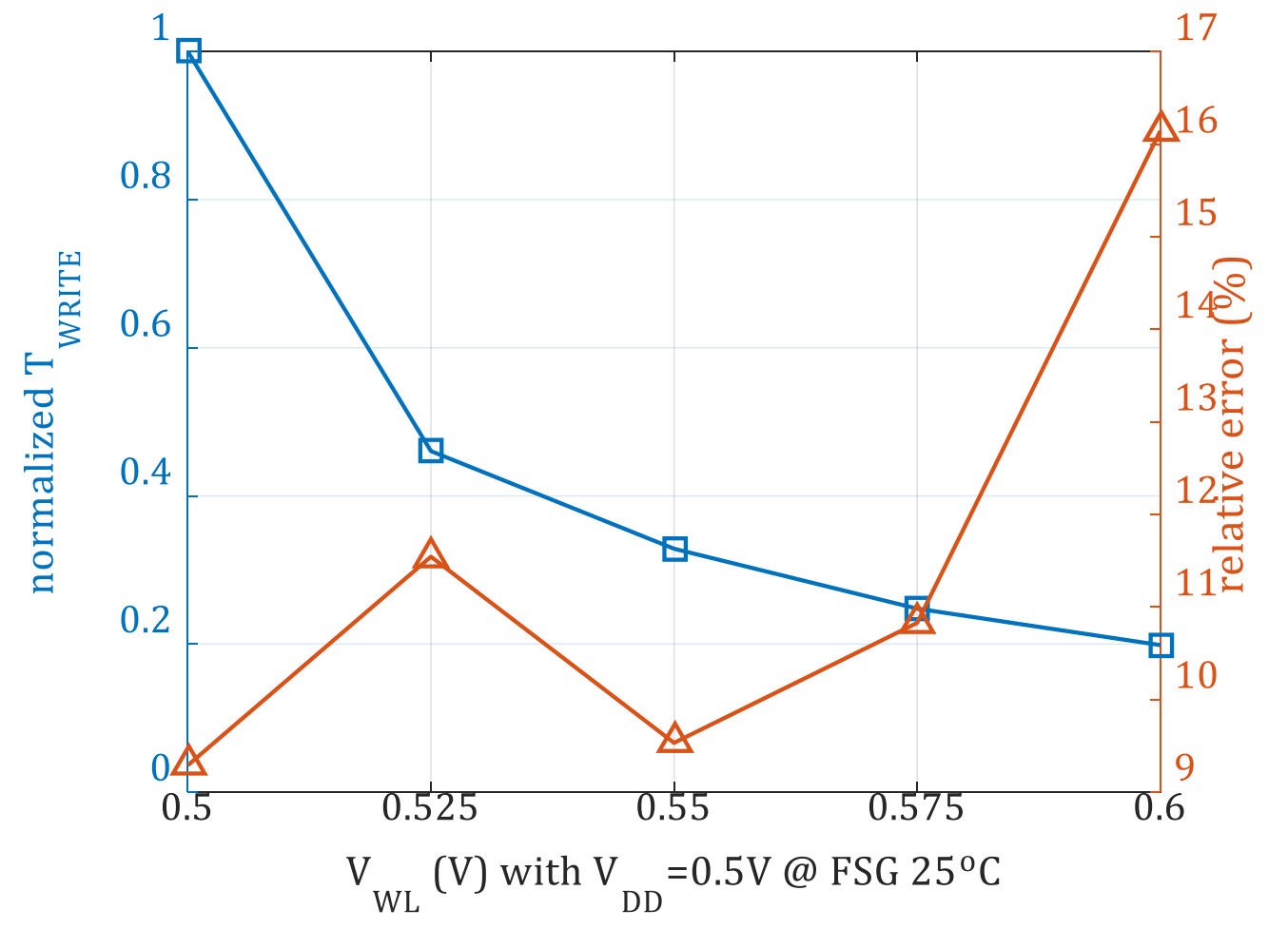

Fig. 11. Normalized $T_{WRITE}$ (left) and relative errors (right) of the proposed model at different WL voltages 25°C FSG corner.

Fig. 11 depicts the normalized $T_{WRITE}$ and relative errors of the write failure model at different WL voltages using WL-boost assist circuits under 4σ constraint. The maximum relative error of our model is 16.1% at 0.6V $V_{WL}$. The assist circuits enhance both the write margin and speed at low supply voltages. A 25mV $V_{WL}$ boost reduces more than half of $T_{WRITE}$ at 0.5V 25°C. However, the write assist technique introduces additional power and area overhead, especially when a large boost capacitor is used [30].

Different from other studies [9][10][11][14], our simulations run with full process parameter variations of all transistors in the netlist. As a consequence, the evaluation results are closer to the real situation in the design flow and more valuable for designers.

To further compare the efficiency of our model, Table IV lists the number of samples required by several importance-sampling methods to evaluate the 4σ failure probability. The number of samples of this work is significantly smaller

TABLE IV
COMPARISONS OF IMPORTANCE SAMPLING METHODS.

| | MC | HDIS [13] | AIS [12] | This work |
|---|---|---|---|---|
| SAMPLES | 1.3E7 | >25500 | >8720 | 200(A)/1600(W) |
| MAX ERROR | - | 10% | 13.9%(A)/ 34.1%(W) | 15.7%(A)/ 17.3%(W) |
| SPEEDUP | 1X | 510X | 1500X | 65000X(A)/ 8125X(W) |

A for access-time failure prediction, W for write failure prediction compared with other sampling-based methods. For each PVT condition, we collect 200 MC samples of $\Delta V$ to calculate $\mu_\Delta$ and $\sigma_\Delta$ in (13) and 200 samples of offset voltages to obtain $\mu_{V_{os}}$ and $\sigma_{V_{os}}$ in (3), where size of data set is 43.6X and 127.5X smaller than that of AIS [12] and HDIS [13] respectively. For predicting the write failure, 1600 $T_{WRITE}$ samples are enough to quantify $\mu_W$ and $\sigma_W$ in (17) and achieve the best accuracy. Furthermore, we find that the number of samples will affect the final evaluation results where larger samples may result in overfitting given a certain form of PDF. The importance-sampling-based methods need a new pre-sampling procedure as PVT changes. Moreover, the failure prediction of AIS is limited to a fixed access time $T_{READ}$, and it needs to be re-constructed when the timing constraint has changed. Our models can evaluate a wide range of failure probabilities since the PDFs of interest metrics have been constructed.

## V. Conclusion

Yield estimation for SRAM cells requires a significantly high range of σ. It is a hard but critical task because the yield results can be widely used by error-tolerant strategies or read/write assist circuit design. Unfortunately, the precision and the speed of evaluation are a trade-off. To tackle this challenge, two analytical failure models are proposed to evaluate the SRAM dynamic stabilities in read and write behaviors. A precise compact drain current model for nanoscale transistors operated at near/sub-threshold voltages is constructed as the foundation of deriving our analytical models. They can offer accurate predictions and physical insights into timing failures.

## Acknowledgment

This work was supported in part by the National Natural Science Foundation of China (NSFC) under Grant 61974024 and Grant 61874152, and the Excellent Doctoral Dissertation Fund of Southeast University under Grant YBPY2023.

## References

[1] Mukhopadhyay, Saibal, Hamid Mahmoodi, and Kaushik Roy. "Modeling of failure probability and statistical design of SRAM array for yield enhancement in nanoscaled CMOS." *IEEE transactions on computer-aided design of integrated circuits and systems* 24, no. 12 (2005): 1859-1880.

[2] Das, Jayita, and Swaroop Ghosh. "Energy barrier model of SRAM for improved energy and error rates." *IEEE Transactions on Circuits and Systems I: Regular Papers* 61, no. 8 (2014): 2299-2308.

[3] Kang, Heechai, Jisu Kim, Hanwool Jeong, Young Hwi Yang, and Seong-Ook Jung. "Architecture-Aware Analytical Yield Model for Read Access in Static Random Access Memory." *IEEE Transactions on Very Large Scale Integration (VLSI) Systems* 23, no. 4 (2014): 752-765.

[4] Jeong, Hanwool, et al. "One-sided static noise margin and Gaussian-Tail-Fitting method for SRAM." *IEEE TVLSI* 22.6 (2013): 1262-1269.

[5] Shen, Shan, et al., “TYMER: A Yield-based Performance Model for Timing-speculation SRAM”, Presented at DAC,2020.

[6] Agarwal, Kanak, and Sani Nassif. "Statistical analysis of SRAM cell stability." In *Proceedings of the 43rd annual Design Automation Conference*, pp. 57-62. 2006.

[7] Fan, Xin, Rui Wang, and Tobias Gemmeke. "Physical modeling of bitcell stability in subthreshold SRAMs for leakage-area optimization under PVT variations." In *Proceedings of the International Conference on Computer-Aided Design*, pp. 1-8. 2018.

[8] Dong, Wei, Peng Li, and Garng M. Huang. "SRAM dynamic stability: Theory, variability and analysis." In *2008 IEEE/ACM International Conference on Computer-Aided Design*, pp. 378-385. IEEE, 2008.

[9] Wang, Joseph, et al. "Non-Gaussian distribution of SRAM read current and design impact to low power memory using voltage acceleration method." *2011 Symp. on VLSI*. IEEE, 2011.

[10] Shiomi, Jun, Tohru Ishihara, and Hidetoshi Onodera. "Microarchitectural-level statistical timing models for near-threshold circuit design." In *The 20th Asia and South Pacific Design Automation Conference*, pp. 87-93. IEEE, 2015.

[11] Cao, Peng, Zhiyuan Liu, Jingjing Guo, and Jiangping Wu. "An Analytical Gate Delay Model in Near/Subthreshold Domain Considering Process Variation." *IEEE Access* 7 (2019): 171515-171524.

[12] Shi, Xiao, Fengyuan Liu, Jun Yang, and Lei He. "A fast and robust failure analysis of memory circuits using adaptive importance sampling method." In *2018 55th ACM/ESDA/IEEE Design Automation Conference (DAC)*, pp. 1-6. IEEE, 2018.

[13] Wu, Wei, Fang Gong, Gengsheng Chen, and Lei He. "A fast and provably bounded failure analysis of memory circuits in high dimensions." In *2014 19th Asia and South Pacific Design Automation Conference (ASP-DAC)*, pp. 424-429. IEEE, 2014.

[14] Shi, Xiao, et al. "Meta-model based high-dimensional yield analysis using low-rank tensor approximation." DAC 2019.

[15] Pang, Liang, Shan Shen, and Mengyun Yao. "A Spline-High Dimensional Model Representation for SRAM Yield Estimation in High Sigma and High Dimensional Scenarios." *IEEE Access* 9 (2021): 47320-47329.

[16] Rabaey, Jan. *Low power design essentials*. Springer Science & Business Media, 2009.

[17] Keller, Sean, David Money Harris, and Alain J. Martin. "A compact transregional model for digital CMOS circuits operating near threshold." *IEEE Transactions on Very Large Scale Integration (VLSI) Systems* 22, no. 10 (2013): 2041-2053.

[18] Van Langevelde, R., and G. Gildenblat. "PSP: An advanced surface-potential-based MOSFET model." In *Transistor Level Modeling for Analog/RF IC Design*, pp. 29-66. Springer, Dordrecht, 2006.

[19] Chang, Tsung-Yung Jonathan, et al. "A 5-nm 135-Mb SRAM in EUV and High-Mobility Channel FinFET Technology With Metal Coupling and Charge-Sharing Write-Assist Circuitry Schemes for High-Density and Low-V MIN Applications." *IEEE Journal of Solid-State Circuits* 56.1 (2020): 179-187.

[20] Shen, Shan, et al. "TS cache: A fast cache with timing-speculation mechanism under low supply voltages." IEEE Transactions on Very Large Scale Integration (VLSI) Systems 28.1 (2019): 252-262.

[21] Yang, Jun, et al. "A Double Sensing Scheme With Selective Bitline Voltage Regulation for Ultralow-Voltage Timing Speculative SRAM." IEEE Journal of Solid-State Circuits 53.8 (2018): 2415-2426.

[22] Khalil, DiaaEldin, et al. "Accurate estimation of SRAM dynamic stability." *IEEE Transactions on Very Large Scale Integration (VLSI) Systems* 16.12 (2008): 1639-1647.

[23] Shen, Shan, et al. "Modeling and Designing of a PVT Auto-tracking Timing-speculative SRAM." *2020 Design, Automation & Test in Europe Conference & Exhibition (DATE)*. IEEE, 2020.

[24] Chang, Montoye, Nakamura, et al. "An 8T-SRAM for Variability Tolerance and Low-Voltage Operation in High-Performance Caches". IEEE Journal of Solid-State Circuits, 2008, 43(4):956-963.

[25] Chang, Ik Joon, et al. "A 32 kb 10T sub-threshold SRAM array with bit-interleaving and differential read scheme in 90 nm CMOS." *IEEE Journal of Solid-State Circuits* 44.2 (2009): 650-658.

[26] Rabaey, Jan M., Anantha P. Chandrakasan, and Borivoje Nikolić. *Digital integrated circuits: a design perspective.* Vol. 7. Upper Saddle River, NJ: Pearson Education, 2003.

[27] Hu, C., X. Xi, M. Dunga, J. He, W. Liu, K. M. Cao, X. Jin, J. Ou, M. Chan, and A. Niknejad. "Bsim4. 5.0 mosfet model." *User's Manual* (2004).

[28] Ohbayashi S, Yabuuchi M, Nii K, et al. A 65-nm SoC embedded 6T-SRAM designed for manufacturability with read and write operation stabilizing circuits[J]. IEEE Journal of Solid-State Circuits, 2007, 42(4): 820-829

[29] Zhang K, Bhattacharya U, Chen Z, et al. A 3-GHz 70-Mb SRAM in 65-nm CMOS technology with integrated column-based dynamic power supply[J]. *IEEE Journal of Solid-State Circuits*, 2006, 41(1): 146-151

[30] Sinangil M E, Mair H, Chandrakasan A P. A 28nm high-density 6T SRAM with optimized peripheral-assist circuits for operation down to 0.6V[C]. In: *IEEE International Solid-State Circuits Conference, Digest of Technical Papers*. 2011. 260-262

[31] Sharma V, Cosemans S, Ashouei M, et al. A 4.4 pJ/Access 80 MHz, 128 kbit variability resilient SRAM with multi-sized sense amplifier redundancy[J]. *IEEE Journal of Solid-State Circuits*, 2011, 46(10): 2416-2430

[32] Lee, Inhak, et al. "24.3 A Voltage and Temperature Tracking SRAM Assist Supporting 740mV Dual-Rail Offset for Low-Power and High-Performance Applications in 7nm EUV FinFET Technology." *2019 IEEE International Solid-State Circuits Conference-(ISSCC)*. IEEE, 2019.

[33] Wu, Jui-Jen, et al. "A Large $\sigma $ V $ _ {\rm TH} $/VDD Tolerant Zigzag 8T SRAM With Area-Efficient Decoupled Differential Sensing and Fast Write-Back Scheme." *IEEE Journal of Solid-State Circuits* 46.4 (2011): 815-827.

[34] Li, T. Bengtsson B., and P. Bickel. "Curse-of-dimensionality revisited: Collapse of importance sampling in very high-dimensional systems." *University of California, Berkeley, Department of Statistics, Tech. Rep* 696 (2005).